\documentclass[11pt]{article}%
\usepackage{amssymb}
\usepackage{amsfonts}
\usepackage{amsmath}
\usepackage[nohead]{geometry}
\usepackage[doublespacing]{setspace}
\usepackage{indentfirst}
\usepackage{graphicx}%
\usepackage{rotating}
\usepackage{float}
\usepackage{color}
\usepackage{booktabs}
\usepackage{pdflscape}
\usepackage{float}
\usepackage[T1]{fontenc}
\usepackage[utf8]{inputenc}
\usepackage{fancyhdr}
\usepackage{natbib}
\usepackage{multirow}
\usepackage{caption}
\usepackage{subcaption}

\usepackage{titling}
\date{}
\predate{}
\postdate{}

\makeatletter
\def\@biblabel#1{\hspace*{-\labelsep}}
\makeatother
\nonstopmode

\begin{document}
\sloppy

\title{\textbf{Testing a Goodwin model with general capital accumulation rate}}
\author{Matheus R. Grasselli\thanks{Corresponding author: grasselli@math.mcmaster.ca, Department of Mathematics and Statistics , McMaster University.
The authors are grateful for comments and suggestions by two anonymous referees, David Harvie, Miguel Leon Ledesma, Steve Keen, Roberto Veneziano, and Alejandro Gonzalez, as well as by the participants of the Post-Keynesian Winter School (Grenoble, December 2015) where part of this work was presented. This research received partial financial support from the Institute for New Economic Thinking (Grant INO13-00011) and the Natural Sciences and Engineering Research Council of Canada (Discovery Grants).} \ 
and
Aditya Maheshwari\thanks{Department of Statistics and Applied Probability, University of California, Santa Barbara.}}
\maketitle

\begin{abstract}
We perform econometric tests on a modified Goodwin model where the capital accumulation rate is constant but not necessarily equal to one as in the original model \citep{Goodwin1967}. In addition to this modification, we find that addressing the methodological and reporting issues in \citet{Harvie2000} leads to remarkably better results, with near perfect agreement between the estimates of equilibrium employment rates and the corresponding empirical averages, as well as significantly improved estimates of equilibrium wage shares. Despite its simplicity and obvious limitations, the performance of the modified Goodwin model implied by our results show that it can be used as a starting point for more sophisticated models for endogenous growth cycles.  
\end{abstract}

\textbf{Keywords:} Goodwin model, endogenous cycles, parameter estimation, employment rate, income shares.

\textbf{JEL Classification Numbers: C13, E11, E32.} 

\section{Introduction}
\label{intro} 

Compared with the large literature dedicated to theoretical aspects of the Goodwin model and its extensions\footnote{The seminal paper 
is \citet{Goodwin1967}. \citet{Desai1973} introduced three extensions of the model to incorporate inflation, expected inflation and variable capacity utilization. In a series of papers \citep{vanderPloeg1984, vanderPloeg1985, vanderPloeg1987}, van der Ploeg introduced the impact of savings by households, substitutability between labor and capital and cost minimizing impact of technical change. There has also been numerous works aimed at understanding the impact of government policy within the framework of the Goodwin model (see \citealp{Wolfstetter1982, TakeuchiYamamura2004, Asada2006, YoshidaAsada2007, CostaLimaGrasselliWangWu2014}). They address problems such as choice of policy (Keynesian versus Classical), role of policy lag and types of debt. Recently \citet{NguyenHuuCostaLima2014} introduced a stochastic version of the Goodwin model with Brownian noise in the productivity factor}, empirical studies 
of such models have attracted relatively little interest. A well-known exception is \citet{Harvie2000}, where the Goodwin model is put to test 
using data for 10 OECD countries from 1959 to 1994, with largely negative conclusions. Regrettably, the work presented in \citet{Harvie2000} contained a serious mistake, 
as well as several smaller problems with its methodology and data construction, that call its conclusion into question. The purpose of the present paper
\footnote{A previous version of this work circulated with the title `Econometric Estimation of Goodwin Growth Models'. The key difference in this version is the capital accumulation rate $k$ introduced in Section \ref{Goodwin_model_setup}. The results obtained in the previous version for the original Goodwin model, namely corresponding to $k=1$, are now presented in an online appendix to this paper, which also contains results for the related Desai and van der Ploeg models and can be found at
www.math.mcmaster.ca/grasselli/appendix-Goodwin} is to address these issues and reevaluate the empirical validity of the Goodwin model. We begin with a brief overview of related empirical studies of the Goodwin and similar models. 

One of the first studies that tried to analyze the Goodwin model in the context of real data was \citet{Atkinson1969}, with an emphasis on 
finding the typical time scale of long-run steady state and cyclical models. Atkinson uses the then recently proposed Goodwin model 
as an example of growth model with cycles and proceeds to calculate its period using several alternative values for the underlying parameters. Although not 
an econometric study, it made attempts to compare periods for trade cycles in postwar United States with those obtained for the Goodwin model. It also inspired the approached later adopted in \citet{Harvie2000}, namely to test the Goodwin model by estimating the underlying 
parameters separately from the model and comparing the resulting equilibrium values (and period) with the corresponding empirical averages. A major breakthrough in the area came with \citet{Desai1984}, where the foundation on how to estimate such dynamic models was laid using data for the United Kingdom for the period 1855 to 1965. By testing generalized models having the Goodwin model as a special case, Desai largely rejected the empirical validity of many assumptions in the Goodwin model. Following Desai, \citet{Harvie2000} tested the Goodwin model in 10 OECD countries from 1959 to 1994 by comparing the estimated equilibrium wage shares and employment rates with the empirical average values. Although he observed qualitative evidence of the cyclical relationship as proposed by Goodwin, there was poor quantitative evidence of the Goodwin model being close to reality. Unfortunately there were several problems with the data construction in this paper, in addition to the mistake described in detail in Section \ref{critique}, that compromised the validity of most of its results. In this regard, \citet{MohunVeneziani2006} have discussed the appropriate data for econometric estimation for the Goodwin model and the problems with Harvie's estimations. Although they did not do econometric estimation, they compared the qualitative cycles and trends for non-farm payroll data for US using two different datasets. Interestingly, they attribute most of the problematic results reported in \citet{Harvie2000}, such as the unrealistically high estimates for the parameters in the Philips curve, to structural change in the data over the period. As we explain in Section \ref{critique}, however, most of these problems disappear once the mistake in \citet{Harvie2000} is corrected.\footnote{For example, the correct parameters in the Philips curve can be obtained by simply dividing the parameters reported in \citet{Harvie2000} by a factor of 100.} In addition, as shown in Section \ref{Goodwin_section_estimate}, we do not find evidence for structural break in the relationships used to estimate the underlying parameters of the model. \citet{GarciaMolinaMedina2010} extended the work done in \citet{Harvie2000} for 67 developed and developing countries. These countries could be divided into three groups, one which depicted Goodwin cycles, the second with movement in opposite direction as predicted by Goodwin due to demand-pushed cycles, and a third without any cyclical movement. 

Other approaches have been used to test the Goodwin model. For example, \citet{Goldstein1997} used multivariate vector auto regression (VAR) specification to understand dynamic interaction between unemployment and profit share of income. He also extended the model to include structural shifts in it. As another example,  \citet{DibehaLuchinskyLuchinskayadSmelyanskiy2007} used the Bayesian inference method to directly estimate the parameters of the differential equations in the Goodwin model, rather than looking at the underlying structural equations like \citet{Harvie2000}. They also modify the classic Goodwin model by introducing a sinusoidal wave to act as exogenous periodic variation. Although the estimates are very close to real data, they do not comment on the theoretical properties such as structural stability of the new stochastic system. \citet{Flaschel09} extended the literature by analyzing the wage share-employment rate relationship using modern econometric techniques. He used a Hodrick-Prescott filter to decompose the state variables into trend and cycles for the US economy. He found considerable evidence of the closed Goodwin cycles that is more prominent than looking at raw data. Further he used a non-parametric bivariate P-spline regression to understand the relationship between wage share and employment rate and the dynamics of unemployment-inflation rate. An important contribution in this study is the separation of long phase cycles and business cycles through this method. \citet{Tarassow2010} used bivariate VAR model for quantifying the relationship between wage share and employment rate in the US economy. The core of the paper is the implementation of impulse response functions and variance decomposition of forecast error to understand the propagation of shocks in one variable to the other using both the raw data and HP filtered data.  
\citet{MassyAvilaGarciaMolina2013} introduced multiple sine-cosine terms to the state equations in the Goodwin model in order to better explain the fluctuations in real data for 16 countries. Although addition of harmonics definitely improved the fit, there is no discussion on the theoretical properties or the impact on structural stability of the model. Recently \citet{MouraRibeiro2013} took a non-conventional route to estimate the Goodwin model, as well as an extension proposed by \citet{DesaiHenryMosleyPemberton2006}, using data for the Brazilian economy from 1981-2009. The novelty of their approach lies in the data construction for wage share and employment rate.  They use the Gompertz-Pareto distribution on individual income database for Brazil to find the wage share and profit share. Moreover since the methodology to calculate unemployment changed over the years, they redefined unemployment as a state when the average individual income is equal to or below 20\% of the national minimum salary. Using these two new data series, they estimate classic Goodwin and its extension. Although there is clear evidence of qualitative cycles, they do not find quantitative evidence to support the Goodwin model or its extension. 

Our own approach is much closer to that of \citet{Desai1984} and \citet{Harvie2000}, but with the aim to, first of all, address the problems in \citet{Harvie2000},  
then extend the study to a broader and more systematic dataset, and finally perform empirical tests of an extension of the Goodwin model. For this, we first update the data used in \citet{Harvie2000} to cover a longer period from 1960 to 2010 and redefine some of the key variables taking into account the criticisms raised, among others, in \citet{MohunVeneziani2006}. Next we introduce what turns out to be a crucial modification in the original Goodwin model, namely allowing the ratio of investment to profit to be given by a parameter $k$, which should then be estimated from the data along with the other parameters in the model, rather than assumed to be identically equal to one as in the original Goodwin model.\footnote{We are indebted to an anonymous referee for suggesting this modification.}

We then perform an empirical test of the Goodwin model in Section \ref{Goodwin Model} along the lines suggested in \citet{Harvie2000}, namely by estimating the underlying parameters of the model and comparing the resulting estimates for the equilibrium values of employment rate and wage share with the corresponding empirical means over the period. This includes a careful analysis of stationarity of the underlying time series and stability of the estimated parameters. We find a marked improvement over the results reported in \citet{Harvie2000}. For example, the estimates for equilibrium employment rate are remarkably close to the empirical 
means, with an average relative error of just 0.53\% across all countries, ranging from a minimum relative error of 
0.1\% in Germany and to a maximum of 1.15\% in Canada and Finland. By comparison, the estimates for equilibrium 
employment rate in \citet{Harvie2000} were not even inside the range of observed data, resulting in an average relative error of 9.09\% 
across all countries. As we mention in Section \ref{critique} and discuss in detail in \citet{GrasselliMaheshwari2017}, most of this improvement in 
the estimated employment rates can be attributed to correcting the reporting mistake in \citet{Harvie2000}. Our results for wage shares, on the other hand, are 
also significantly better than those of \citet{Harvie2000}, even though they were not affected by the same mistake. The improvement in this case is largely attributable to the introduction of the capital accumulation rate $k$, which \citet{Harvie2000} implicitly assumes to be equal to one, in accordance with the original Goodwin model, but we estimate from the data. As a result, our estimated equilibrium wage shares lie well within the range of observed values for all countries and have an average relative error of 2.54\% when compared to empirical means, ranging from a minimum relative error of 0.26\% for Germany to a maximum of 5.83\% for the UK. By comparison, the estimated equilibrium wage shares estimated in \citet{Harvie2000} for the original Goodwin model are outside the range of observed value for all countries and have an average relative error of 38\%, ranging from a minimum relative error of 22.6\% for Norway to a maximum of 103\% for Greece.\footnote{As explained in Section \ref{data}, we use the same number of countries as \citet{Harvie2000}, but replace Greece with Denmark, whose economic 
fundamentals are closer to the other countries in the sample. Excluding Greece, the average relative error in the estimated equilibrium wage share in \citet{Harvie2000} is still 30.8\%.}

More importantly, the introduction of the capital accumulation rate $k$ also leads to improved performance for the simulated trajectories of the modified Goodwin model obtained from the estimated parameters. We show this in Section \ref{simulated} by means of the Theil statistics, where we compute the root-mean-square errors between for 
employment rates and wage shares using all observed points and the corresponding simulated trajectories. We find that the errors are again smaller for employment rates than for wages shares, which also show a larger component of systematic errors.

\section{A modified Goodwin model}
\label{Goodwin Model}

This section explains theoretical setup of the original Goodwin model as proposed in \citet{Goodwin1967} and the modification adopted in this paper. This is followed by  an explanation of the corresponding econometric setup presented in \citet{Harvie2000} and a description of the data and summary statistics. 

\subsection{Model Setup}
\label{Goodwin_model_setup}

The Goodwin model starts by assuming a Leontieff production function with full capital utilization, that is,
\begin{equation}
Y(t)=\min\left\{\frac{K(t)}{\nu},a(t)L(t)\right\} \label{leontiffProductionFn}%
\end{equation}
where $Y$ is real output, $K$ is real capital stock, $L$ is the employed labor force, $a$ is labor productivity and $\nu$ is a constant capital-to-output ratio. It also assumes exponential growth function for both productivity and total labour force of the form\begin{align}
N(t) &= N(0)e^{\beta t} \\
a(t) & =a(0)e^{\alpha t},
\end{align} 
where $\alpha$ and $\beta$ are constant growth rates. Finally, the version of the Goodwin model adopted in this paper is based on
the following two behavioural relationships:
\begin{align}
\label{Philips_continuous}
\frac{\dot{\mathrm{w}}}{\mathrm{w}} &= \Phi (\lambda) = \gamma + \rho \lambda  \\
\dot{K} &= \Pi-\delta K = k(Y-\mathrm{w}L) -\delta K,
\label{investment}
\end{align}
where $\gamma,\rho,k$ and $\delta$ are constants.  
The first relationship above says that the growth in real wage rate 
\begin{equation}
\mathrm{w(t)}=\frac{W(t)}{L(t)},
\label{wage_rate}
\end{equation}
where $W$ denotes the real wage bill in the economy, depends 
on the employment rate
\begin{equation}
\label{employment_rate}
\lambda(t)=\frac{L(t)}{N(t)}
\end{equation} through a linear Phillips curve $\Phi$. The second relationship, namely equation \eqref{investment} above, says that a constant fraction $k$ of total 
profits 
\begin{equation}
\Pi(t)=Y(t)-W(t)
\label{profits}
\end{equation}
from production are reinvested in the accumulation of capital, which in turn depreciates 
at a constant rate $\delta$. The remainder fraction $(1-k)$ of profits are distributed as dividends to the household sector, which is assumed 
to have no savings, so that all wages and dividends are spent on consumption. 

\begin{figure}[H]
   \centering  
     \includegraphics[width=0.9 \textwidth]{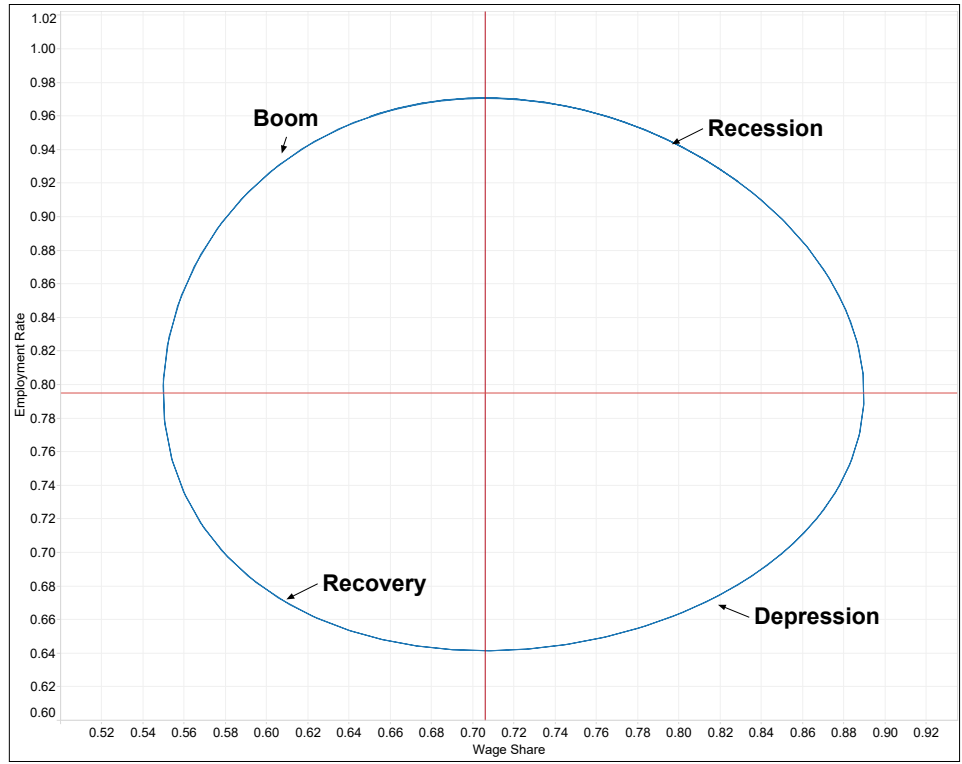} 
   \caption{Solution for the Goodwin model \eqref{goodwin_model_1}-\eqref{goodwin_model_2} with 
parameter values $\alpha=0.018$, $\beta=0.02$, $\delta=0.06 $, $\gamma=0.3 $, $\rho=0.4 $, $\nu=3$, $k=1$.}
   \label{Goodwin_fig} 
\end{figure}

Using these assumptions and defining $\omega=\frac{wL}{Y}$ as the wage share of output in the economy, one can derive the following set of equations to describe the relationship of wage share and employment rate:
\begin{align}
\label{goodwin_model_1}
\frac{\dot{\omega}}{\omega} &=\gamma + \rho \lambda-\alpha \\
\frac{\dot{\lambda}}{\lambda} &=\frac{k(1-\omega)}{\nu}-(\alpha+\beta+\delta).
\label{goodwin_model_2}
\end{align} 
The solution of these system of differential equations is a closed orbit around the non-hyperbolic equilibrium point 
\begin{align}
\label{lambda_equi}
\overline{\lambda} & = \frac{\alpha-\gamma}{\rho}\\
\overline{\omega} & = 1-(\alpha+\beta+\delta)\frac{\nu}{k},
\label{omega_equi}
\end{align}
with period given by 
\begin{equation}
\label{period}
T=\frac{2\pi}{\left[(\alpha-\gamma)(k/\nu-(\alpha+\beta+\delta))\right]^{1/2}},
\end{equation}
and is illustrated in Figure \ref{Goodwin_fig}. In the original Goodwin model proposed in \citet{Goodwin1967}, investment was assumed to be always equal to profits, that is to say, $k=1$ in \eqref{investment}. To the best of our knowledge, the more general form in \eqref{investment}, with a constant $k$ not necessarily 
equal one, was first proposed in \citet{Ryzhenkov2009} in the context of a more complicated three-dimensional model for the wage share, employment rate, and a variable capital-to-output ratio. 

\begin{center} [ Insert Figure \ref{Goodwin_fig} here ]
\end{center}

\subsection{Econometric setup}
\label{critique}

The test of the Goodwin model proposed by \citet{Harvie2000} consists of comparing the {\em econometric-estimate predictors} for the equilibrium point $(\overline{\lambda},\overline{\omega})$, which can be obtained from 
\eqref{lambda_equi}-\eqref{omega_equi} by substituting the econometric estimates for the underlying parameters in the model, with the 
empirical average of the observed employment rates and wage shares through the data sample.

Before describing our results, we take a slight detour to discuss some of the methodological and reporting issues in \citet{Harvie2000}. To start with, \citet{Harvie2000} had a transcription mistake in the reported estimated parameters for the Phillips curve: the coefficients for the employment rate in Table A2.3 of \citet{Harvie2000} are incorrect.\footnote{We thank David Harvie for informing us through private communication that the coefficients of the employment rate were supposed to be in percentages but mistakenly used as numbers. For example, the estimated coefficient for the employment rate for Australia is reported as $-86.73$ in Table A2.3 when it should 
have been $-0.8673$.} This mistake propagated further, leading to inappropriate equilibrium estimates of employment rate in Table 2 of \citet{Harvie2000}. The mistakenly large estimates of the parameters in the Phillips curve effectively killed the impact of productivity growth on employment rate and led to estimates of employment rate that were downward biased, with over 10\% absolute error for some countries. For example, if the correct coefficients from Table A2.3 had been used, the estimate for the equilibrium employment rate for UK would have been 0.96, while \citet{Harvie2000} reported it as 0.85. The estimate of period of the Goodwin cycle is also incorrect due to same problem and consequent miscalculation. The correct period of the Goodwin cycles should have been between 10 to 22 years for the data used in \citet{Harvie2000}, but it was reported to be between one and two years for all the countries. The mistake and its consequences, as well as the correct values 
for the parameters, equilibrium points, and periods are discussed in detail in \citet{GrasselliMaheshwari2017}.

\begin{figure}[H]
   \centering  
     \includegraphics[width=0.9 \textwidth]{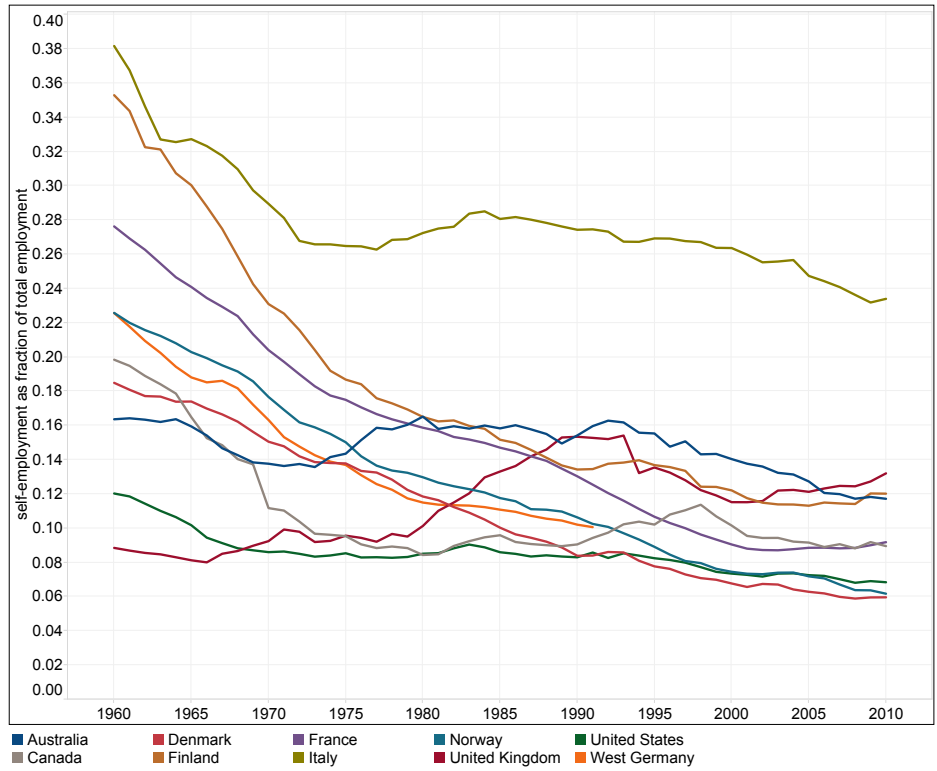} 
   \caption{Self-employment as fraction of total employment. Source: AMECO database.}
   \label{selfEmploymentFig} 
\end{figure}

Secondly, the definition of wage share in his study does not segregate the income of the self employed into labor and capital income. Including proprietor's income as part of `net operating surplus' is a gross underestimation of the wage share. As can be seen in Figure \ref{selfEmploymentFig}, the fraction of the labor force that was self employed in the sample countries during the period of study can be quite high. As of 2010, while Italy has around a quarter of its population as self employed, three of the other 8 countries have over 10\% of total employment as self employed. This effect was more prominent in early part of the data, when 8 of the 10 countries had over 15\% of total employment as self employed. Including their total income as part of profits is therefore inappropriate. 

\begin{center} [ Insert Figure \ref{selfEmploymentFig} here ]
\end{center}

Finally, the methodology in \citet{Harvie2000} was inconsistent in defining the total income of the economy. When defining wage share, \citet{Harvie2000} used (Compensation of Employees + Net Operating Surplus) as a proxy for total income, leaving out consumption of fixed capital and taxes on production and imports from the GDP, whereas when defining productivity and in the derivation of equilibrium values it used GDP as a proxy for total income. In the results that follow, we address these problems with the methodology in \citet{Harvie2000}.

\subsection{Data Construction and Sources}
\label{data}
We use data from the AMECO database\footnote{AMECO is the annual macro-economic database of the European Commission's Directorate General for Economic and Financial Affairs (DG ECFIN). We used the data provided in tabular form at http://knoema.com/ECAMECODB2014Mar/annual-macro-economic-database-march-2014} from 1960 to 2010 for Australia, Canada, Denmark, Finland, France, Italy, Norway, UK and US. For Germany, 
we use data from 1960 to 1990 only, to avoid dealing with the jumps that occur in all variables because of unification. These are the same countries analyzed in 
\citet{Harvie2000}, except that we replaced Greece by Denmark, which has economic fundamentals that are closer to the other countries in the sample. 

We define output as GDP at factor cost, that is, net of taxes and subsidies on production and imports, and use a GDP deflator to obtain real income as
\begin{equation}
Y=\frac{\text{GDP at current prices - net taxes on production and imports}}{\text{GDP Deflator}}.
\end{equation}
This is because the Goodwin model does not consider either taxes or subsidies explicitly, which are included in measures of GDP at producers's price. 

For the estimation of the Goodwin model, we have to separate income into wages and profits. Wages can be gauged from the `Compensation of Employees' variable in the database, but this does not include the income of the self employed, which can be significantly high. Since we can not find segregation of proprietors income into labor and capital, we follow \citet{KlumpMcAdamWillman2007} and use compensation per employee as proxy for labor income of the self employed. Thus the real wage bill in the economy is given by 
\begin{equation}
W=\left(1+\frac{\text{Self Employed}}{\text{Total Employees}}\right)\times\frac{\text{Compensation of Employees}}{\text{GDP Deflator}}
\end{equation}
and gross real profits are defined as $\Pi=Y-W$. We next define total employment as
\begin{equation}
L = \text{total employees + self employed}
\end{equation}
and total labor force as
\begin{equation}
N = \text{total employment + total unemployed}.
\end{equation}
For variables using real capital stock $K$, we use the total net capital stock (at 2005 prices) from the database, which include both private and government fixed assets,  and divide it by real output $Y$ to obtain the capital-to-output 
ratio $\nu=K/Y$. Similarly, the return on capital $r$ can then be defined as 
\begin{equation}
\label{return_rate}
r=\frac{\Pi}{K}.
\end{equation}
For depreciation rate $\delta$, we use the definition from the manual of AMECO database, that is, 
\begin{equation}
\delta=\frac{\text{Consumption of fixed capital at current prices}}{\text{Price deflator for gross fixed capital formation}*\text{Net capital stock (2005 Prices)}}.
\label{depreciation}
\end{equation}
Similarly, for the accumulation rate $k$ we use 
\begin{equation}
k=\frac{\text{gross capital formation}}{\Pi}.
\label{depreciation}
\end{equation}

\subsection{Summary Statistics}

Table \ref{summaryStats} summarizes the data for wage share $\omega=\frac{W}{Y}$ and employment rate $\lambda=\frac{L}{N}$ for the 10 countries we analyze. The average wage share for the period varied between 61.48\% for Norway to 71.47\% for UK, whereas the average employment rate varied between 92.80\% for Italy to 97.31\% for Norway. Norway is a curious case with highest average employment rate (and least variability) and lowest average wage share (and highest variability). Finland is the only country with over 4\% standard deviation in the employment rate, mostly attributable to a slump in the Finnish economy during the early 1990s, when the employment rate dropped by over 12\% in four years and the wage share continued to decline throughout the decade. On the other hand, the United States has one of the most stable wage share and employment rate when compared with any of its European counterpart. 

\begin{center} [ Insert Table \ref{summaryStats} here ]
\end{center}

\section{Estimation results}
\label{Goodwin_section_estimate}

The estimate of the equilibrium employment rate $\overline{\lambda}$ in equation \eqref{lambda_equi} depends on the estimation of Phillips curve $\Phi$, that is 
to say the parameters $\gamma$ and $\rho$, and the productivity growth rate $\alpha$, whereas the estimate of the equilibrium wage share 
$\overline{\omega}$ in equation \eqref{omega_equi} depends on $\alpha$, $\beta$, $\delta$, $k$ and $\nu$. We can estimate the parameters for productivity growth rate and population growth rate using the log-regression of the variables on the time trend, that is, 
\begin{align}
\log(a_t) &= \log(a_0)+\alpha t+\epsilon_{1t} \label{productvtyeqn}\\
\log(N_t) &= \log(N_0)+\beta t+\epsilon_{2t} \label{labourgrowtheqn}
\end{align}

Table \ref{productivityGrowth} presents the estimates of the parameters in equation \eqref{productvtyeqn} for different countries. 
The productivity growth rate $\alpha$ varies from 1.3\% for Canada to 2.9\% for Finland, with all the European countries exhibiting a higher 
productivity growth rates than the three non-European ones. Similarly, Table \ref{laborGrowth} shows the parameters estimate for equation \eqref{labourgrowtheqn}. Here Australia and Canada top the list with roughly 2\% growth rate of labor force followed by US with 1.65\%. All the European economies considered, except Norway, face the problem of ageing population with the growth rate less than 1\%.  Figure \ref{productvtylabourGraph} also gives a graphical interpretation of labour force growth and productivity growth rate.

The estimate for the Phillips curve is more involved. Following \citet{Harvie2000}, we first approximate the term $\dot{\mathrm{w}}/\mathrm{w}$ by 
$(\mathrm{w}_t-\mathrm{w}_{t-1})/\mathrm{w}_{t-1}$ and replace \eqref{Philips_continuous} with
\begin{equation}
\label{Philips_discrete}
z_t = \gamma + \rho \lambda_t,
\end{equation}
for the discrete-time variables $\lambda_t$ and 
\begin{equation}
z_t=\log(\mathrm{w}_t)-\log(\mathrm{w}_{t-1}),
\end{equation}
which is itself an approximation for $(\mathrm{w}_t-\mathrm{w}_{t-1})/\mathrm{w}_{t-1}$. \citet{Harvie2000} then proposes to estimate an autoregressive distributed lag (ARDL) 
model of the form 
\begin{equation}
\label{ARDL}
z_t = a_{0} + a_{1} z_{t-1} + \ldots a_{p} z_{t-p}+ b_0\lambda_{t}+b_{1}\lambda_{t-1} + \ldots b_{q}\lambda_{t-q} + \epsilon_{t},
\end{equation}
and assumes stationarity of the variables to obtain estimates $\widehat\gamma$ and $\widehat\rho$ for the long-run coefficients from the estimates of the ARDL(p,q) model 
above (see \citet[footnote 1, page 356]{Harvie2000}). This is problematic, since there is no guarantee that the variables at hand are indeed stationary. Table \ref{unitRoot} shows the results of the  augmented Dickey-Fuller test (ADF test) to check for unit root for real wage growth, employment rate, productivity growth, inflation and nominal wage growth for the 10 countries in the study. At a broad level we can say that real wage growth and productivity growth are stationary whereas the employment rate, inflation and nominal wage growth are non-stationary for most of the countries.

Because real wage growth and employment rate have different order of integration (the former is stationary, the latter is not), we can not use standard time series models to estimate the parameters in \eqref{Philips_discrete}. Instead, we shall use the bounds-testing procedure proposed by \citet{PesaranShinSmith2001}, which allows us to test for the existence of linear long-run relationship when variables have different order of integration. We start by formulating an unrestricted error correction model (ECM) of the form
\begin{equation}
\Delta z_t = \phi_0 +  \sum\limits_{i=1}^p \phi_{1i} \Delta z_{t-i} + \phi_1\Delta \lambda_{t-1} + \phi_2 z_{t-1} + \phi_3 \lambda_{t-1} + \epsilon_{1t}
\end{equation}
where the lag $p$ is determined using a Bayesian Information Criterion. As it happens, the optimal lag turns out to be zero for all countries, so that the effective 
unrestricted error correction model  is given by 
\begin{equation}
\Delta z_t = \phi_0 +  \phi_1\Delta \lambda_{t-1} + \phi_2 z_{t-1} + \phi_3 \lambda_{t-1} + \epsilon_{1t}.
\label{uecm1}
\end{equation}
We first perform a Ljung-Box Q test to check for no serial correlation in the errors for equation \eqref{uecm1}, as this is a necessary condition for the 
bounds-testing procedure of \citet{PesaranShinSmith2001} to apply.  We observe in Table \ref{serialCorrel_modela} that the p-values for the alternative hypothesis that 
the errors are AR(m) for $m=1,\ldots,5$ are greater than 10\% for all countries, thus implying no serial correlation.

We proceed by testing the hypothesis $H_0:\phi_2 = \phi_3 = 0$ in \eqref{uecm1} against 
the alternative hypothesis that $H_0$ is not true. We do this because, as in conventional co-integration tests, the absence of a long-run equilibrium 
relationship between the variables $z_t$ and $\lambda_t$ is equivalent to $H_0$, so a rejection of $H_0$ implies a long-run relationship. The technical complication 
associated with an arbitrary mix of stationary and non-stationary variables is that exact critical values for a conventional F-test are not available in this case. The 
essence of the approach proposed in \citet{PesaranShinSmith2001} consisted in providing bounds on the critical values for the asymptotic distribution of the F-statistic instead, 
with the lower bounds corresponding to the case where all variables are $I(0)$ and the upper bound corresponding to the case where all variables are $I(1)$. 
The lower and upper bounds provided in \citet{Narayan2005} for 50 observations at 1\%, 5\% and 10\% levels are (7.560, 8.685), (5.220, 6.070) and (4.190, 4.940), respectively. Table \ref{ftest_modela} shows the computed F-statistic for the joint restriction $\phi_2 = \phi_3 = 0$, which lie above the upper bound at the 1\% significance level for all the countries except Germany, where it is above the upper bound at the 10\% significance level. We thus reject the null hypothesis of absence of co-integration for all countries.

Having established that the variables show co-integration, we can now meaningfully estimate a long-run ``levels model'' of the form 
\begin{equation}
z_t = \gamma + \rho \lambda_t + \epsilon_{2t}.
\label{ltpc}
\end{equation}
Table \ref{linearPC} shows the estimates for \eqref{ltpc} where we can see that all countries have negative intercept and positive slope. Thus employment has significantly positive impact at 10\% level of significance for all the countries except Canada, with the coefficient $\rho$ ranging from 11\% for Canada to 75\% for Germany. Italy has a higher coefficient 98\% but it may be plagued by bias due to auto-correlation in the errors. We also perform a test of 
serial correlation in the long-run model \eqref{ltpc}. The last five columns in Table \ref{linearPC} show the p-values for the alternative hypothesis that 
the errors are AR(m) for $m=1,\ldots,5$ and suggests that the model is appropriate for all the countries except Italy, where errors are serially correlated at the 5\% level of significance.

As a final check, we estimate the coefficients of a restricted error correction model of the form 
\begin{equation}
\Delta z_t = \phi_{10} +  \phi_{11}\Delta \lambda_{t-1} + \phi_{12}v_{t-1} + \epsilon_{3t},
\label{recm1}
\end{equation}
where $v$ is a conventional ``error-correction term'' obtained as the estimated residual series from the long-run relationship \eqref{ltpc}, that is,
\begin{equation}
\label{residual}
v_{t-1} = z_{t-1}-\widehat\gamma -\widehat\rho\lambda_{t-1},
\end{equation}
where $\widehat\gamma$ and $\widehat\rho$ are the estimated coefficients in \eqref{ltpc}. If the model is correct, the coefficient of the lagged error terms should be negative and significant, as can be seen in Table \ref{RestrictedECM}. Model diagnostic tests for no-autocorrelation and homoscedasticity are accepted for all countries except Italy.  

Before computing the equilibrium values arising from the estimated parameters, we check for structural change in the underlying relationship by testing the null 
hypothesis that the regression coefficients in equations \eqref{uecm1} and \eqref{ltpc} are constant over time. Figure \ref{cusum_model1} shows the result of 
the CUSUM (cumulative sum of recursive residuals) and CUSUMSQ (cumulative sum of recursive squared residuals) tests for the coefficients of \eqref{uecm1}. We can see fluctuations well within the 99\% confidence interval for all countries for the CUSUM test and for all countries except France and Denmark (note that the tests for Germany have different bands because of the smaller number of observations). Very similar results are shown in Figure \ref{cusum_longterm_model1} for the same tests for the coefficients of the long-run model in equation \eqref{ltpc}. We therefore accept the null hypothesis of constant parameters in equations 
 \eqref{uecm1} and \eqref{ltpc} throughout the period.

Having ruled out structural breaks in the underlying relationship, we follow \citet{Harvie2000} and obtain the econometric-estimate predictors for the Goodwin equilibrium values and period by substituting these parameter estimates into equations \eqref{lambda_equi}-\eqref{period}, that is,
\begin{align}
\label{lambda_equi_estimate}
\lambda_G & = \frac{\widehat\alpha-\widehat\gamma}{\widehat\rho}\\
\label{omega_equi_estimate}
\omega_G & = 1-(\widehat\alpha+\widehat\beta+\widehat\delta)\frac{\widehat\nu}{\widehat k} \\
\label{period_estimate}
T_G & = \frac{2\pi}{\left[(\widehat\alpha-\widehat\gamma)(\widehat k/\widehat\nu-(\widehat\alpha+\widehat\beta+\widehat\delta))\right]^{1/2}}.
\end{align}
These equilibrium estimates for the Goodwin model are presented in Table \ref{parameterEstimates}. In this table, the parameter estimates $\widehat\alpha$ and 
$\widehat\beta$ for the productivity and population growth rate are taken from Tables \ref{productivityGrowth} and \ref{laborGrowth} respectively. The estimate for the depreciation parameter $\widehat\delta$ is the average of the historical depreciation calculated using equation \eqref{depreciation}. Similarly, the estimate for 
capital-to-output ratio $\widehat\nu$ and capital accumulation rate $\widehat k$ are the historical averages of the ratios of real capital stock to real output and 
real investment to real profits, respectively. The estimates $\widehat\gamma$ and 
$\widehat\rho$ for the parameters of the linear Phillips curve are taken from Table \ref{linearPC}. 

\begin{center} [ Insert Table \ref{parameterEstimates} here ]
\end{center} 

\begin{figure}[H]
     \centering  
     \includegraphics[width=0.8 \textwidth]{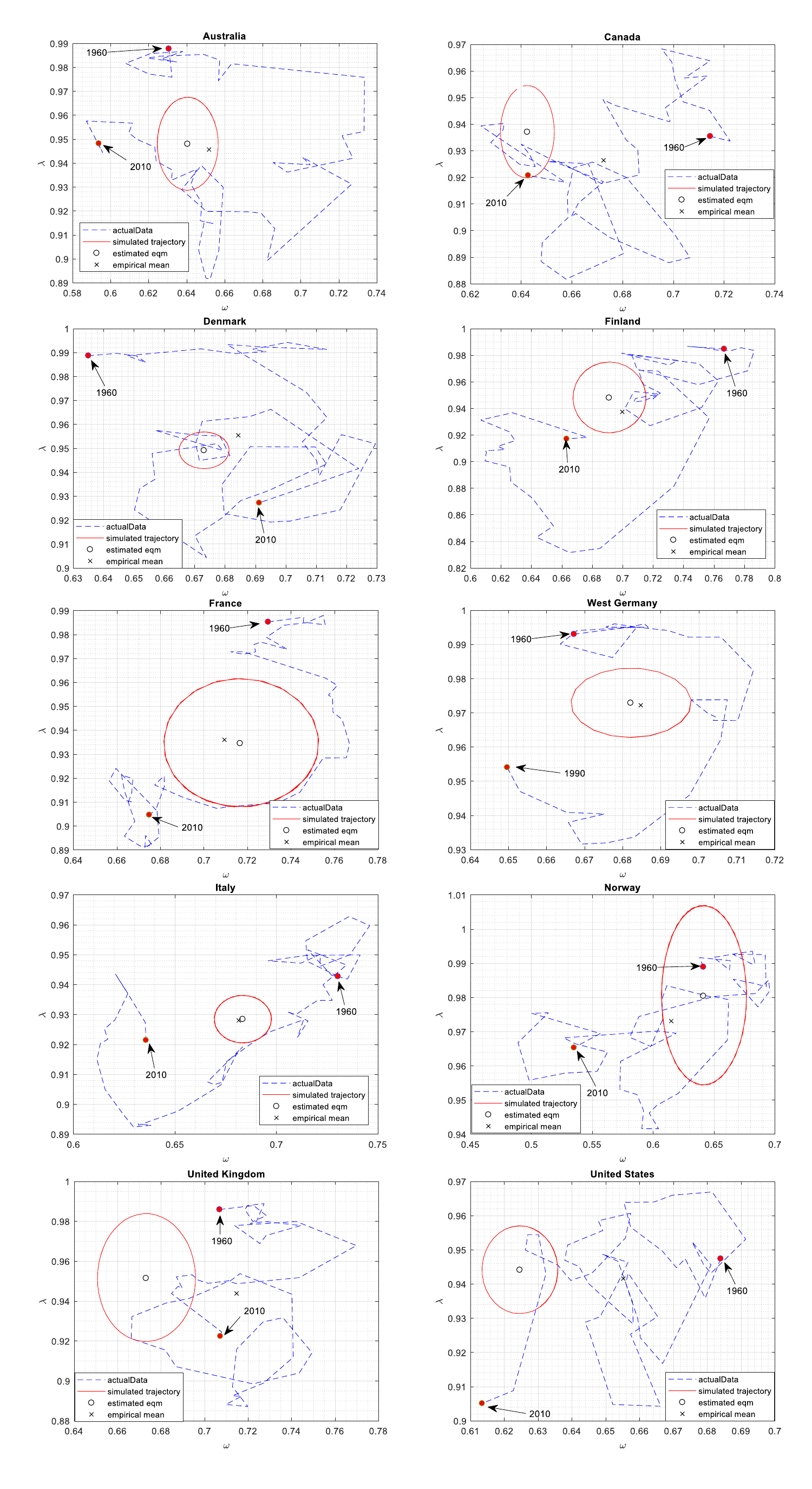} 
  \caption{Observed data and corresponding empirical mean for wage share and employment rates, together with estimated equilibrium points and simulated trajectories for the 
   modified Goodwin model.}
   \label{finalGraphs} 
\end{figure}

As we can see in Figure \ref{finalGraphs}, the estimates for the equilibrium wage share $\omega_G$ and employment rate $\lambda_G$ are well within the range of observed values. Table \ref{error_comparison} shows the absolute and relative errors for the estimated when compared with the corresponding empirical means over the period. Starting with the employment rate, we see that the absolute difference $|\overline \lambda-\lambda_G|$ between the empirical mean $\overline \lambda$ reported in Table \ref{summaryStats} and the 
estimated equilibrium value $\lambda_G$ reported in Table \ref{parameterEstimates} is less than 1\% for all the countries except Canada and Finland, where the differences are 1.07\% and 1.05\%, respectively. The relative error $|\overline \lambda-\lambda_G|/\overline \lambda$ for the employment rate ranges from 0.05\% for Italy to 1.15\% for Canada and averages to 0.53\% over the countries in the sample. Compared with the estimated values in \citet{Harvie2000}, which were not even inside the range of observed data and had an average relative error of 9.09\%, this is a motivating improvement. As mentioned in Section \ref{intro}, the reason for the high errors in the estimates for equilibrium employment rates reported in  \citet{Harvie2000} was the transcription mistake explained in Section \ref{critique}. When correcting for this mistake, as shown in \citet{GrasselliMaheshwari2017}
the average relative error in equilibrium employment rate is reduced from 9.09\% to 0.60\%, which is comparable with the average relative error of 0.53\% obtained here. 

Moving on to the wage share, we see from Table \ref{error_comparison} that the absolute difference $|\overline \omega-\omega_G|$ between the empirical mean $\overline \omega$ reported in Table \ref{summaryStats} and the estimated equilibrium value $\omega_G$ reported in Table \ref{parameterEstimates} is less than 3\% for all the countries except the UK and the US, where the differences are 4.2\% and 3.1\%, respectively. The relative error $|\overline \omega-\omega_G|/\overline \omega$ for the wage share ranges from 0.26\% for Germany to 5.83\% for the UK and averages to 2.54\% over the countries in the sample. This is a remarkable improvement in performance when compared with the estimated values in \citet{Harvie2000}, which were outside the range of observed data and had an average relative error of nearly 40\%, ranging from 22\% for the UK to more than 100\% for Greece. Even excluding Greece, which is not part of our dataset, the average relative error for the estimated wage share in \citet{Harvie2000} is more than 30\%. The improvement in estimates for equilibrium wage share, however, have nothing to do with the transcription mistake in \citet{Harvie2000}, since the parameters affected by the mistake only enter in the calculation of the equilibrium employment rate. The improved estimates can be attributed instead to two different factors: (i) a more accurate measurement of the wage share that takes into account self-employment as explained in Section \ref{data} and (ii) the introduction of the investment-to-output ratio $k$ in \eqref{investment}. As can be seen in expression \eqref{omega_equi_estimate}, a lower estimate $\widehat k$ leads to a lower equilibrium wage share. We see from Table \ref{parameterEstimates} that the estimates $\widehat k$ are significantly lower one, which is the value implicitly assumed in the original Goodwin model analyzed in \citet{Harvie2000}. 

\begin{center} [ Insert Figure \ref{finalGraphs} here ]
\end{center}

\section{Simulated Trajectories} 
\label{simulated}

Our approach thus far has concentrated on the measure of performance suggested in \citet{Harvie2000} for the Goodwin model, namely the comparison between 
the estimated equilibrium values for wage share and employment rate and their corresponding empirical means for the period in the sample. An alternative measure consists of analyzing the errors in the actual trajectories, rather than equilibrium values only. In other words, we can simulate the trajectories of the modified Goodwin model implied by the estimated parameters in Table \ref{parameterEstimates} and compute the difference between each observed wage share and employment rate pair  and the corresponding pair on the simulated trajectory. Since there is a closed orbit associated with each initial condition, we repeat this procedure using each observed data pair as a candidate initial condition. For each country, we then select the initial condition that minimizes the mean squared error. Finally, we decompose this mean squared error in order to better understand the sources of error. The results are presented in Table \ref{mse}.  

\begin{center} [ Insert Table \ref{mse} here ]
\end{center}

\begin{figure}[H]
   \centering  
     \includegraphics[width=0.8 \textwidth]{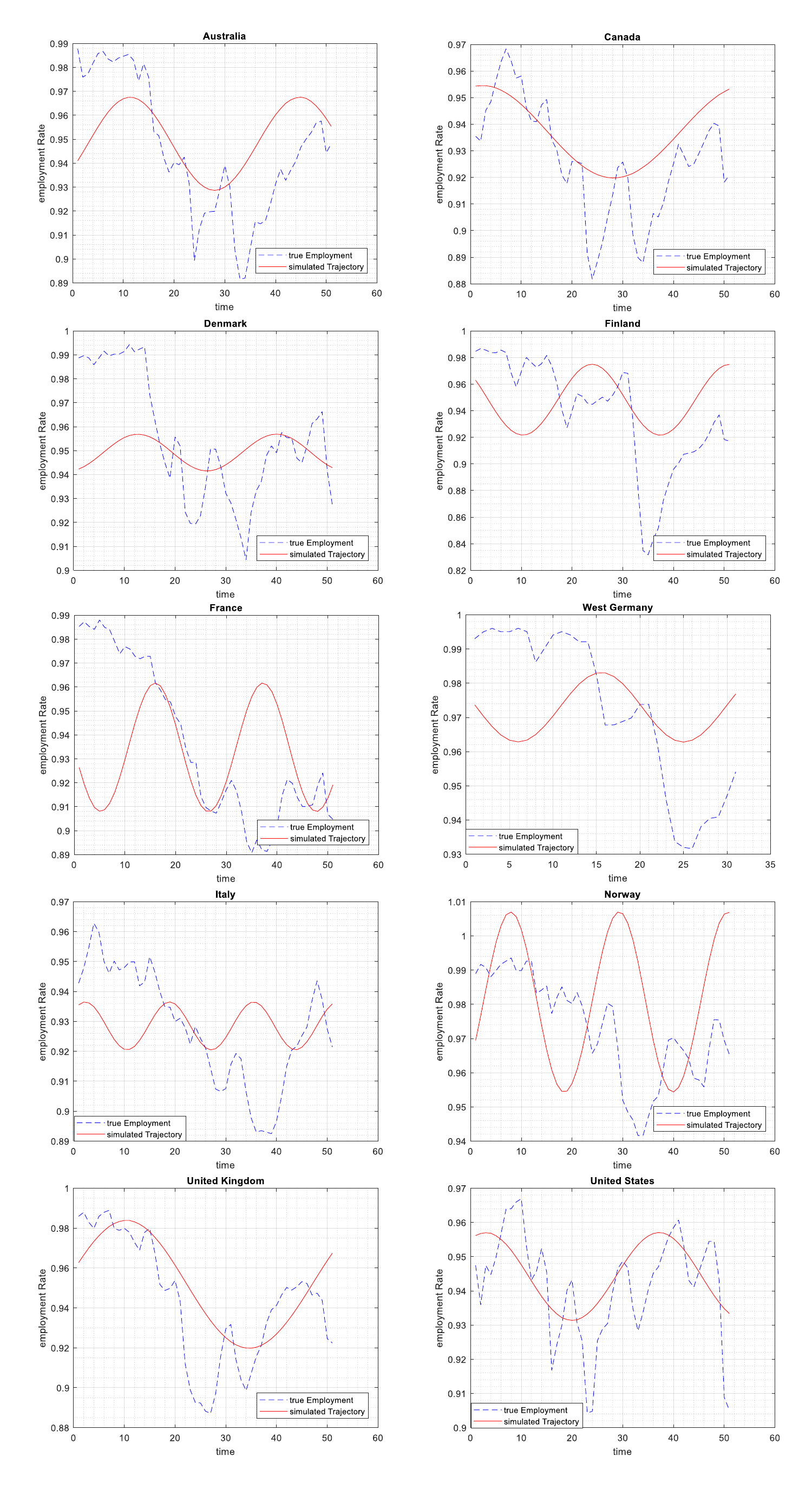} 
   \caption{Observed and simulated employment rates for the modified Goodwin model.}
   \label{employmentGraphs} 
\end{figure}

The first column of Table \ref{mse} shows the root-mean-square error (RMSE) for employment rate as a fraction of the mean employment rate over the period. We see that this ranges from 1.4\% for the US to 4.5\% for Finland, with an average of 2.6\% across all countries. The next three columns shows the decomposition of the mean squared error (MSE) into a bias, variance, and covariance proportions. The bias proportion $U^M_\lambda$ indicates how far the mean of the simulated trajectory is from the mean of the observed data, whereas the variance proportion $U^S_\lambda$ indicates how far the variance of the simulated trajectory is from the variance of the observed data. Together, they measure the proportion of the MSE that is attributed to systematic errors. Accordingly, the covariance proportion $U^C_\lambda$ measures the remaining unsystematic errors. We see from Table \ref{mse} that the bias proportion $U^M_\lambda$ and the variance proportion $U^S_\lambda$ contribute on average to 9.5\% and 29\% of the MSE for employment rate, respectively, so that the covariance proportion $U^C_\lambda$ is the largest one and contributes on average to 61.5\% of the MSE. This is a positive result, but masks large differences between the countries. For example, whereas France is 
a model case where both the bias (1.5\%) and the variance (11.7\%) are low, there are examples such as Canada, with a high bias (38.2\%) and low 
variance (9.8\%) contributions and other such as Germany, with very low bias (0.2\%) but high variance (51.8\%) contributions. These differences can be seen in Figure \ref{employmentGraphs}, which shows both the observed data and simulated trajectories for the modified Goodwin model. 

\begin{figure}[H]
   \centering  
     \includegraphics[width=0.8 \textwidth]{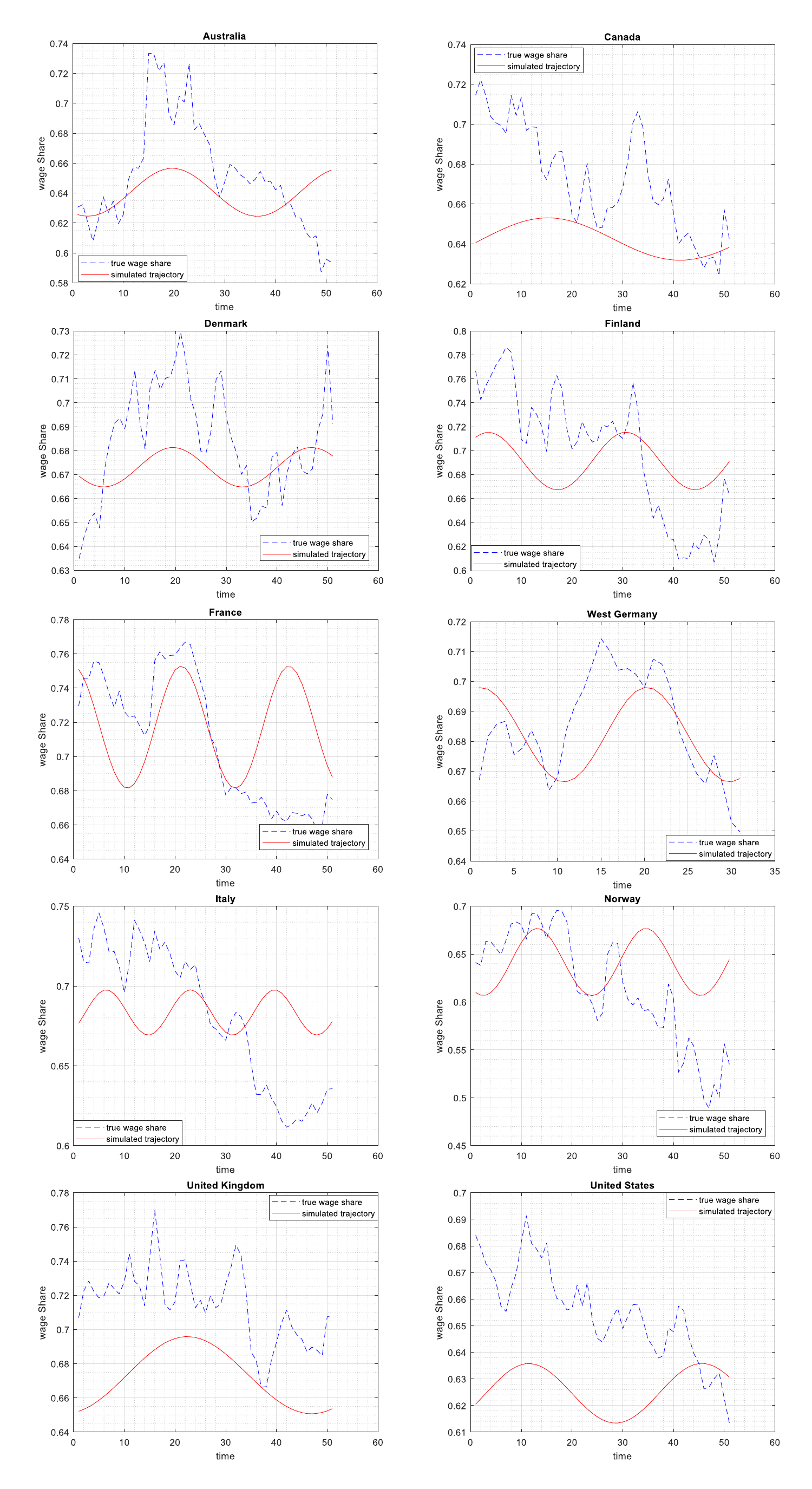} 
 \caption{Observed and simulated wage shares for the modified Goodwin model.}
   \label{wageGraphs} 
\end{figure}

The last four columns in Table \ref{mse} show the corresponding results for the wage share. Consistently with the results in Section \ref{Goodwin_section_estimate}, where we found 
that the errors in equilibrium wage share were systematically higher than the ones for employment rate, we see that the RMSE for the simulated wage share as a fraction of the mean wage share for the period is also higher than the ones for employment rate, averaging at 5.8\% over all countries. We also see that the average bias (27.9\%) and variance (31\%) contributions for the MSE in wage share are higher than the corresponding proportions for the employment rate. In other words, not only the MSE are higher for wage share than for employment rate, but they contain a much larger proportion of systematic error. This can be seen in Figure \ref{wageGraphs}, where the agreement between observed and simulated values for wage shares is generally worse than that for employment rate shown in Figure \ref{employmentGraphs}. In particular, the Goodwin model is clearly unable to match the decreasing trend in wage share observed in most countries, most notably the US, even though it captures the cyclical fluctuations reasonably well. 

\begin{center} [ Insert Figures \ref{employmentGraphs} and \ref{wageGraphs} here ]
\end{center}

\section{Concluding remarks}
\label{conclusion}

The Goodwin model is a popular gateway to a large literature on endogenous growth cycles, as it serves as the starting point to much more complex models, such 
as the model proposed in \citet{Keen1995} and its many extensions. Any hope of empirical validation of the extended models, therefore, necessarily needs to be based 
on a somewhat decent performance of the basic model. The tests performed in \citet{Harvie2000}, however, seemed to have dealt these endeavours a fatal blow by showing 
that the basic Goodwin model was not remotely descriptive of the cycles observed in real data for OECD countries in the second half of the last century. 

The main contribution of this paper is to dispense once and for all with this notion. We show that a simple modification of the Goodwin model, namely the introduction of a parameter
$0<k \leq 1$ representing a constant capital accumulation rate, leads to remarkable improvements in performance when compared with the results reported in \citet{Harvie2000}. In particular, the estimates for $k$ show that it is generally much smaller than one, which corresponds to the implicit assumption in the original Goodwin model. Since a lower value for $k$ leads to a lower equilibrium wage share, our estimates for equilibrium wage share are systematically lower than the ones in \citet{Harvie2000} and much closer to the empirical means. 

We move beyond a simple comparison between equilibrium values and empirical means and analyze the performance of the simulated trajectories for the modified Goodwin mode. We find that both the simulated employment rates and wage shares lie comfortably within the range of observable values, with the single exception of the simulated wage shares for the UK, which lie below the observed values for the entire period. Moreover, the simulated trajectories are not too far from observed values. For example, the root-mean-square errors for employment rates ranges from 1.4\% (US) to 4.5\% (Finland) of the mean employment rate, whereas the root-mean-square errors for wage shares ranges from 2.3\% (Germany) to 9.3\% (Norway) of the mean wage share. Furthermore, we observe that the contribution of unsystematic errors to the mean squared error is on average much larger for employment rates (61.5\%) than for wage shares 
(41.1\%). 

Nevertheless, even in the modified Goodwin analyzed here has clear and severe limitations. As it is quite apparent, the patterns for observed data shown in Figure \ref{finalGraphs} do not even remotely resemble the closed orbits predicted by the model, even though the quantitative errors are not as bad as previously believed. In other words, the model is unable to capture more complicated dynamics for employment rates and wage shares, such as the sub-cycles that can be seen for many countries, or the clear downward trend for wage share. 

Our results suggest, however, that endogenous growth cycle models based on extensions of the Goodwin model deserve much more empirical explorations. In particular, models incorporating more realistic banking and financial sectors, such as the extension proposed in \citet{Keen1995} and analyzed in \citet{GrasselliCostaLima2012} have the potential to improve the estimates of the equilibrium wage share even further, given the more flexible investment behaviour assumed for firms. In addition, models exhibiting a larger variety of dynamic behaviour, such as limit cycles or multiple equilibria, might provide even more accurate descriptions of the type of economic variables treated here.

\bibliographystyle{cjeFinal}
\bibliography{finance}

\newpage 

\appendix

\section{Auxiliary Tables}

\begin{center} [ Insert Tables \ref{productivityGrowth} to \ref{RestrictedECM} here ]
\end{center} 

\section{Auxiliary Figures}

\begin{figure}[H]
\centering
\begin{subfigure}{.9\textwidth}
  \centering
  \includegraphics[width=.9\linewidth]{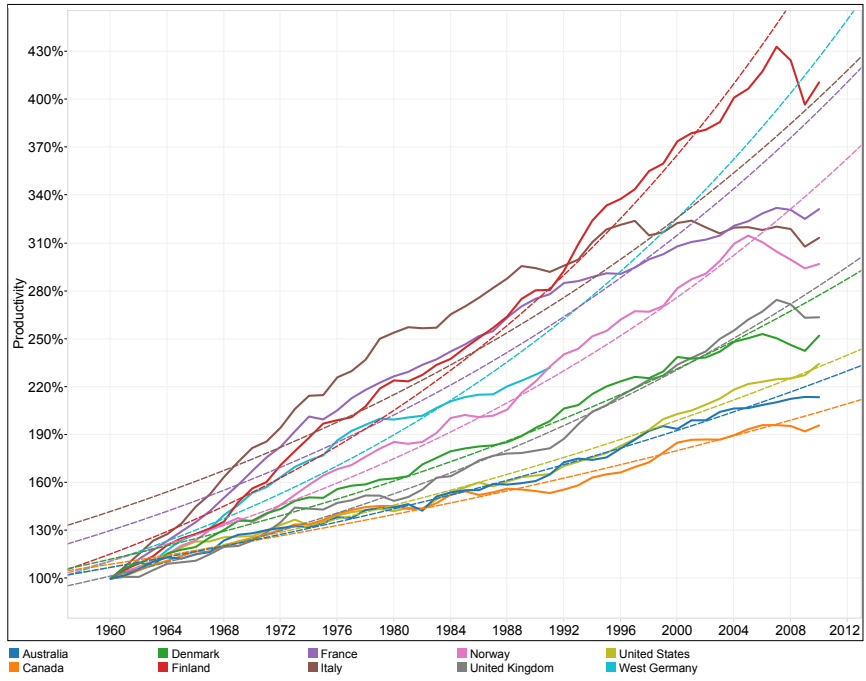}
  \subcaption{Productivity}
  \label{fig:sub1}
\end{subfigure}
\begin{subfigure}{.9\textwidth}
  \centering
  \includegraphics[width=.9\linewidth]{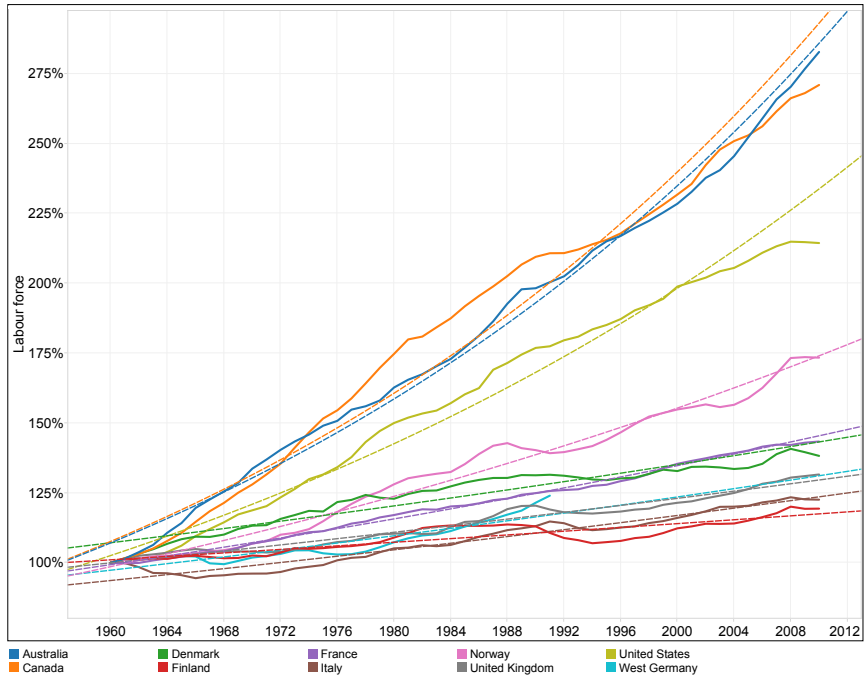}
  \subcaption{Labour Force}
  \label{fig:sub2}
\end{subfigure}
\caption[Productivity and Labour Force]{Both productivity and labour force are presented as proportion of their value in 1960. The dotted lines are the exponential trend lines}
\label{productvtylabourGraph}
\end{figure}

\begin{figure}[H]
      \centering  
     \includegraphics[width=.85 \textwidth]{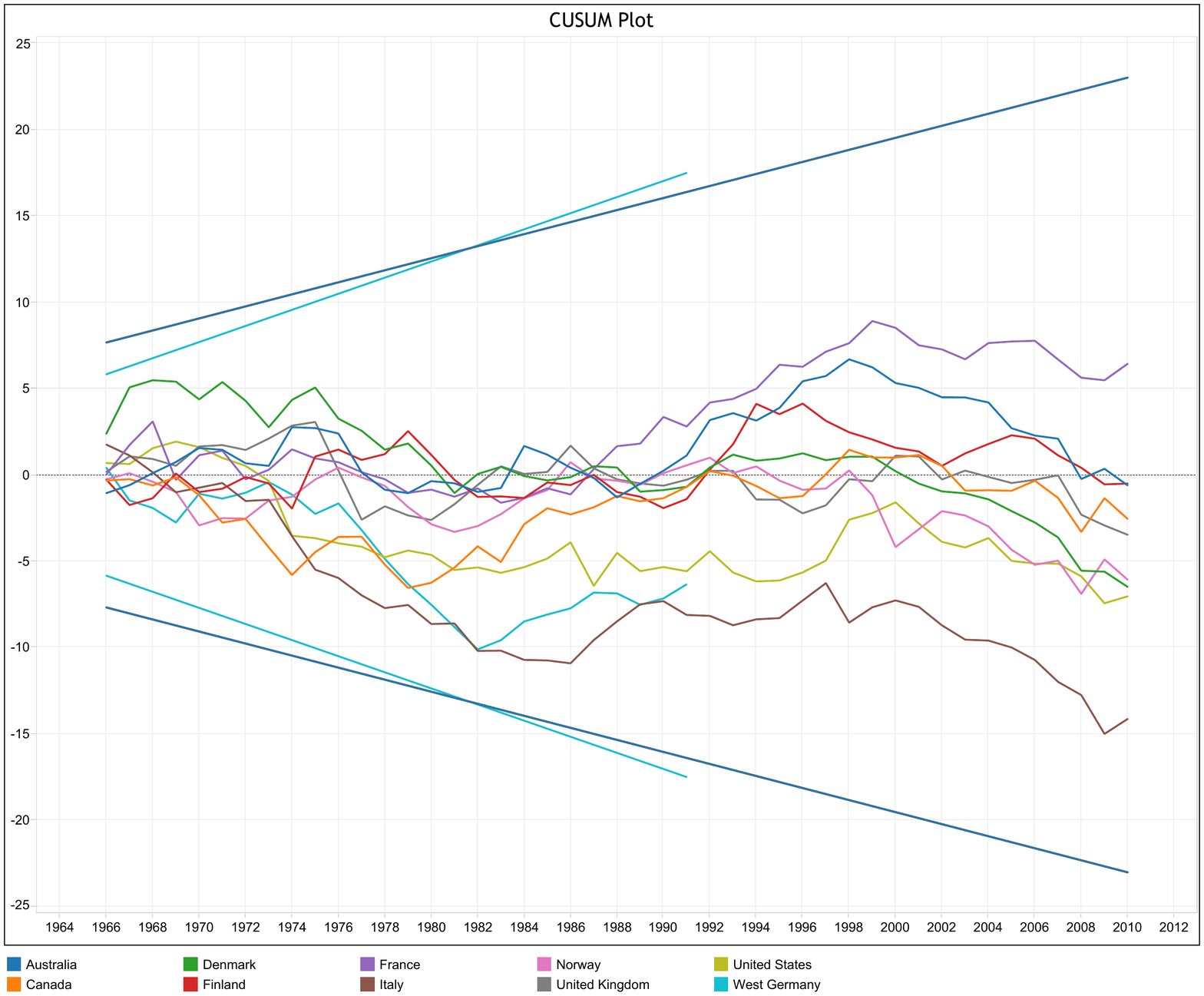} 
     \includegraphics[width=.85 \textwidth]{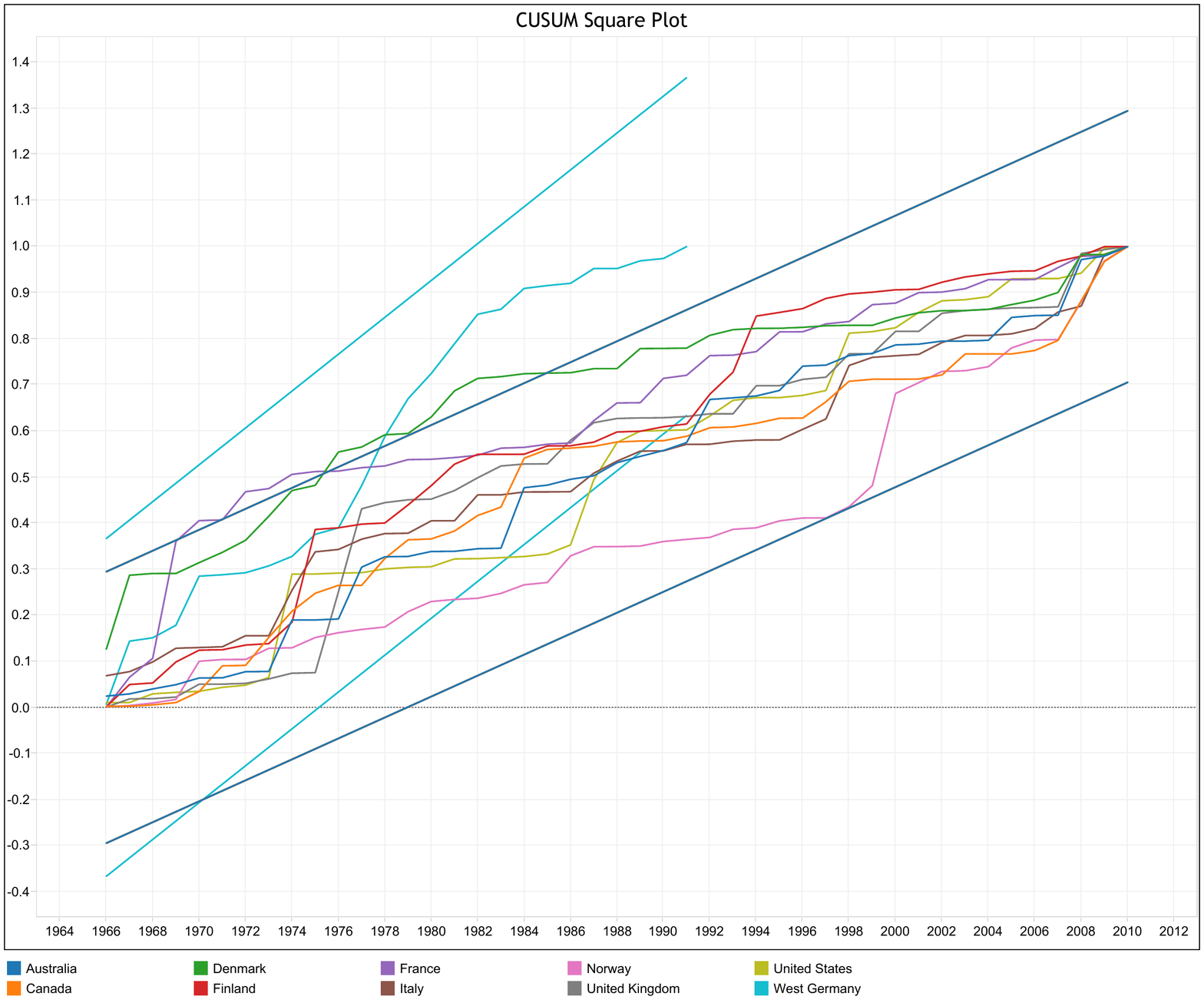} 
    \footnotesize
\caption{Tests for structural changes in \eqref{uecm1} using CUSUM and CUSUMSQ tests at the 99\% confidence interval.}
   \label{cusum_model1} 
\end{figure}

\begin{figure}[H]
   \centering  
     \includegraphics[width=.85 \textwidth]{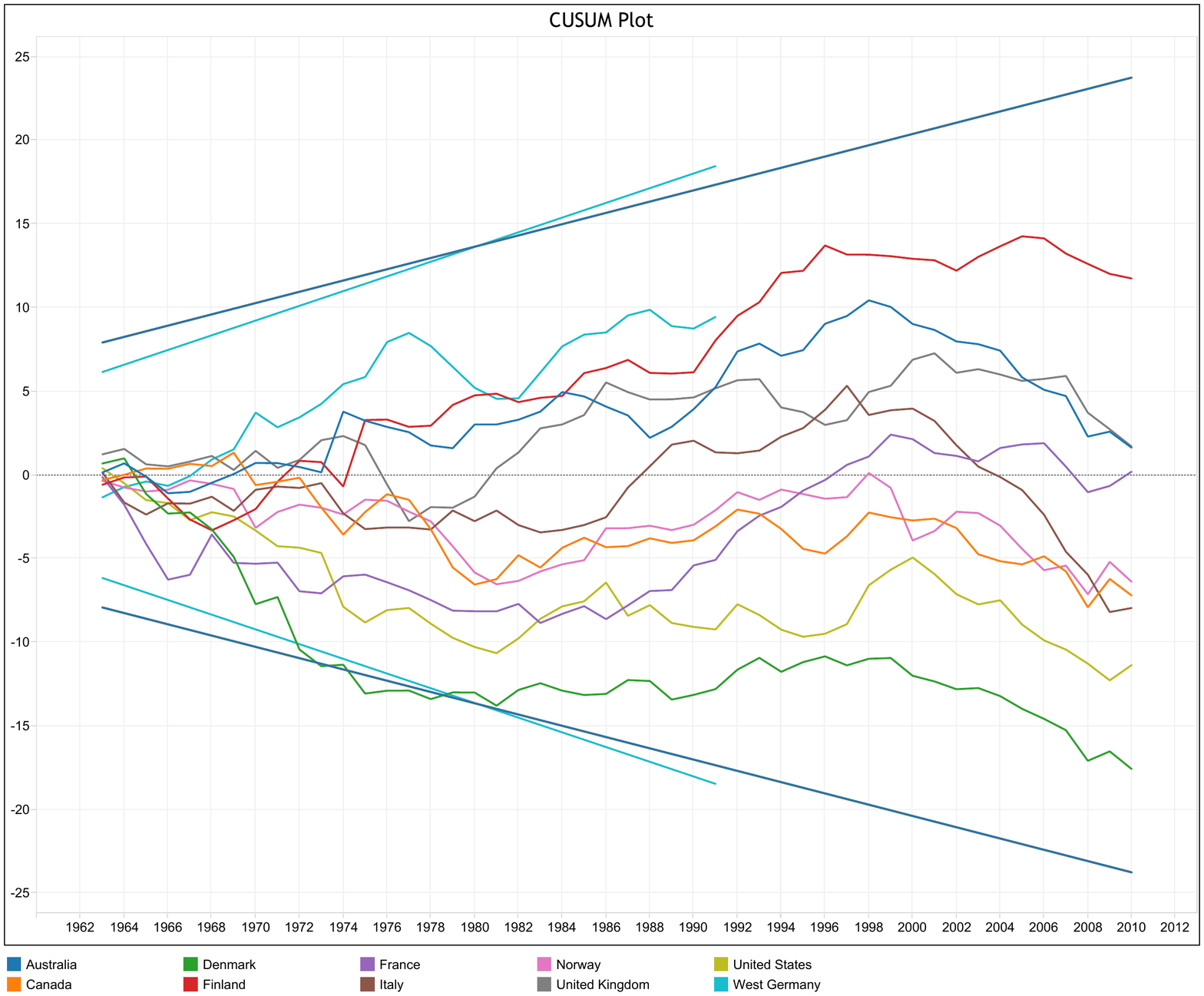} 
     \includegraphics[width=.85 \textwidth]{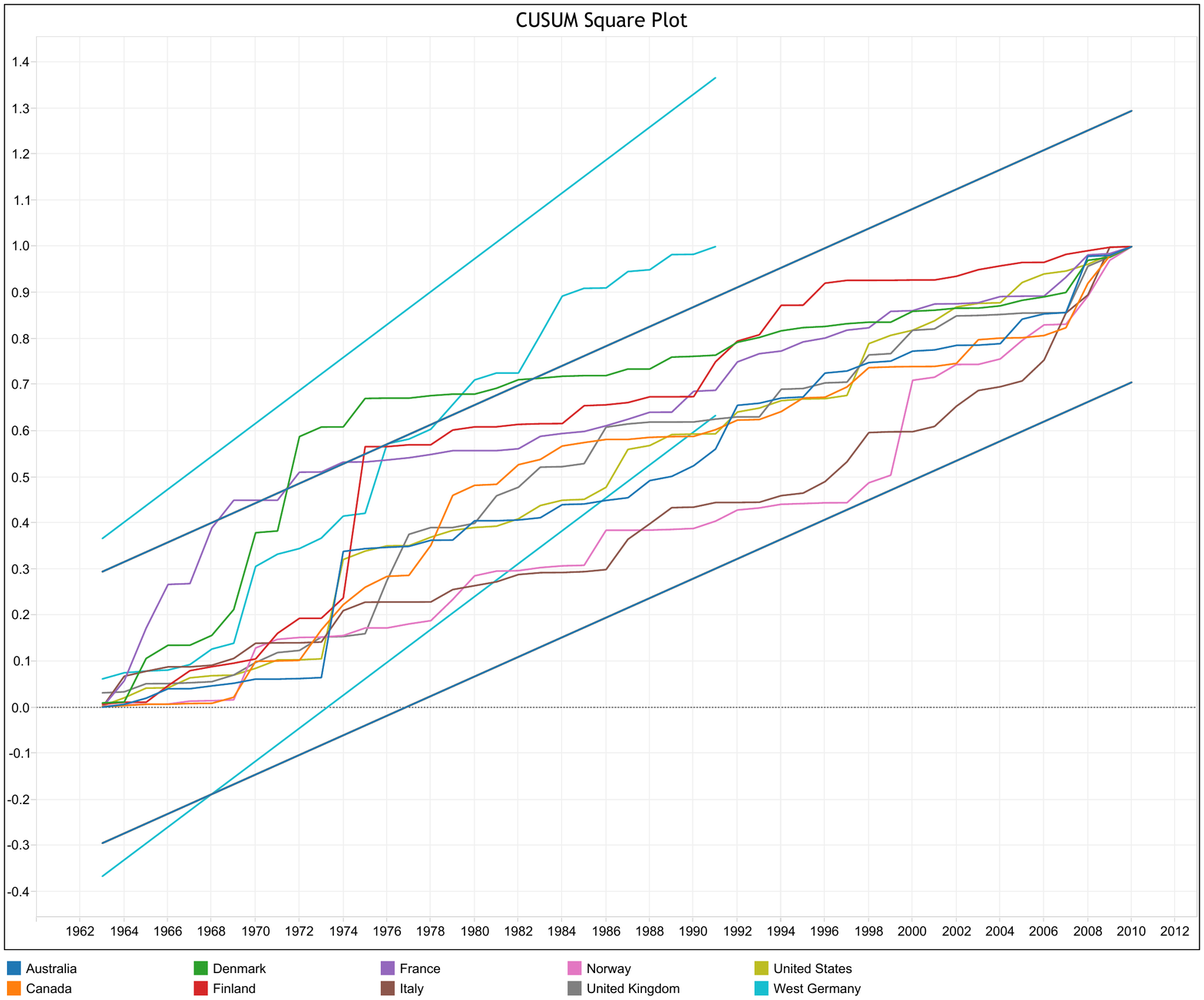} 
    \footnotesize
   \caption{Tests for structural changes in \eqref{ltpc} using CUSUM and CUSUMSQ tests at the 99\% confidence interval.}
   \label{cusum_longterm_model1} 
\end{figure}

\begin{center} [ Insert Figures \ref{productvtylabourGraph} to \ref{cusum_longterm_model1} here ]
\end{center} 

\newpage

\begin{table}[H]
\centering
\begin{tabular}{|c|c|c|c|c|}
\hline
 & \multicolumn{2}{c|}{Wage Share} & \multicolumn{2}{c|}{Employment Rate} \\ \hline
Country & mean & std & mean & std \\ \hline
Australia & 0.6517 & 0.0366 & 0.9457 & 0.0282 \\ \hline
Canada & 0.6724 & 0.0268 & 0.9264 & 0.0214 \\ \hline
Denmark & 0.6843 & 0.0228 & 0.9554 & 0.0258 \\ \hline
Finland & 0.6997 & 0.0539 & 0.9375 & 0.0418 \\ \hline
France & 0.7094 & 0.0379 & 0.9361 & 0.0329 \\ \hline
Germany & 0.6838 & 0.0180 & 0.9719 & 0.0230 \\ \hline
Italy & 0.6814 & 0.0440 & 0.9280 & 0.0193 \\ \hline
Norway & 0.6148 & 0.0592 & 0.9731 & 0.0151 \\ \hline
United Kingdom & 0.7147 & 0.0215 & 0.9438 & 0.0311 \\ \hline
United States & 0.6552 & 0.0172 & 0.9416 & 0.0155 \\ \hline
\end{tabular}
\caption{Summary Statistics - 1960 to 2010.}
\label{summaryStats}
\end{table}

\begin{table}[H]
\centering
\begin{tabular}{|l|c|c|c|c|c|c|c|c|c|c|}
\hline
Country 	& $\widehat\alpha$ 	& $\widehat\beta$ 	& $\widehat\delta$ 	& $\widehat\nu$ 	& $\widehat\gamma$	& $\widehat\rho$ 	& $\widehat k$		& $\omega_G$ 	& $\lambda_G$ 	& $T_G$ \\ \hline
Australia 	& 0.015 			& 0.020 			& 0.052			& 2.881 			& -0.215 				& 0.242 			& 0.694			& 0.6404 		 	& 0.9480 			& 33.41 \\ \hline
Canada 	& 0.013 			& 0.020 			& 0.043 			& 2.864 			& -0.095 				& 0.115 			& 0.605			& 0.6424 			& 0.9371 			& 51.94 \\ \hline
Denmark 	& 0.018 			& 0.006 			& 0.050 			& 2.842 			& -0.330 				& 0.367 			& 0.640			& 0.6730 			& 0.9492 			& 27.34 \\ \hline
Finland 	& 0.029 			& 0.003 			& 0.052 			& 3.314 			& -0.258 				& 0.303 			& 0.898			& 0.6910 			& 0.9480 			& 27.10 \\ \hline
France 	& 0.022 			& 0.008 			& 0.038 			& 3.326 			& -0.491 				& 0.549 			& 0.792			& 0.7165 			& 0.9346 			& 21.25 \\ \hline
Germany 	& 0.028 			& 0.006 			& 0.036 			& 3.367 			& -0.705 				& 0.753 			& 0.735			& 0.6821 			& 0.9729	        		&19.03 \\ \hline
Italy 		& 0.021 			& 0.006 			& 0.047 			& 3.206 			& -0.891 				& 0.982 			& 0.738			& 0.6833 			& 0.9285 			& 16.59 \\ \hline
Norway 	& 0.023 			& 0.011			& 0.047 			& 3.208 			& -0.574 				& 0.609 			& 0.722			& 0.6411 			& 0.9804 			& 21.41 \\ \hline
UK 		& 0.021 			& 0.005 			& 0.037 			& 3.053 			& -0.108 				& 0.135 			& 0.588			& 0.6731 			& 0.9515 			& 48.73 \\ \hline
US 		& 0.016 			& 0.016 			& 0.052 			& 2.725 			& -0.227 				& 0.257 			& 0.610			& 0.6245 			& 0.9441 			& 34.13 \\ \hline
\end{tabular}
\caption{Parameter estimates and equilibrium values for the modified Goodwin model.}
\label{parameterEstimates}
\end{table}

\begin{table}[H]
\centering
\begin{tabular}{|c|c|c|c|c|}
\hline
& \multicolumn{2}{c|}{Employment rate} & \multicolumn{2}{c|}{Wage share}   \\ \hline
 & $|\overline\lambda-\lambda_G|$ & $\frac{|\overline\lambda-\lambda_G|}{\overline\lambda}$ & $|\overline\omega-\omega_G|$ & $\frac{|\overline\omega-\omega_G|}{\overline\omega}$ \\ \hline
Australia 			& 0.0023 & 0.24\%	& 0.011	& 1.74\%     \\ \hline
Canada			& 0.0107 & 1.15\% 	& 0.030 	& 4.47\%	   \\ \hline
Denmark			& 0.0062 & 0.65\% 	& 0.011	& 1.65\% 	    \\ \hline
Finland 			& 0.0105 & 1.12\% 	& 0.009	& 1.24\% 	    \\ \hline
France 			& 0.0015 & 0.16\% 	& 0.007	& 1.01\% 	    \\ \hline
Germany 			& 0.0010 & 0.10\% 	& 0.002	& 0.26\% 	    \\  \hline
Italy 				& 0.0005 & 0.05\% 	& 0.002 	& 0.28\% 	     \\ \hline
Norway 			& 0.0073 & 0.75\% 	& 0.026 	& 4.28\% 	     \\ \hline
United Kingdom	& 0.0077 & 0.82\% 	& 0.042 	& 5.83\% 	     \\ \hline
United States 		& 0.0025 & 0.27\% 	& 0.031 	& 4.68\% 	      \\ \hline
Average 			& 0.0050 & 0.53\% 	& 0.017	& 2.54\%	     \\ \hline
\end{tabular}
\caption{Comparison between empirical means and equilibrium values estimates for employment rate and wage share in the modified Goodwin model  - 1960 to 2010.}
\label{error_comparison}
\end{table}

\begin{table}[H]
\centering
\begin{tabular}{|c|c|c|c|c|c|c|c|c|}
\hline
& \multicolumn{4}{c|}{Employment rate} & \multicolumn{4}{c|}{Wage share}   \\ \hline
 & $\frac{\sqrt{MSE_\lambda}}{\overline\lambda}$ & $U_\lambda^M$ & $U_\lambda^S$ & $U_\lambda^C$ & $\frac{\sqrt{MSE_\omega}}{\overline\omega}$ 
 & $U_\omega^M$ & $U_\omega^S$ & $U_\omega^C$\\ \hline
Australia 			& 0.025 & 6.6\%	& 39.2\% 	& 54.3\% & 0.055 & 12.8\%	& 50.4\%	& 36.8\%     \\ \hline
Canada			& 0.018 & 38.2\%	& 9.8\% 	& 52.0\% & 0.059 & 61.6\%	& 18.9\%	& 19.4\% 	   \\ \hline
Denmark			& 0.026 & 5.2\%	& 65.7\% 	& 29.1\% & 0.034 & 23.9\%	& 50.8\%	& 25.3\%  	    \\ \hline
Finland 			& 0.045 & 4.6\%	& 29.1\% 	& 66.4\% & 0.069 & 3.8\%		& 57.7\%	& 38.5\% 	    \\ \hline
France 			& 0.042 & 1.5\%	& 11.7\% 	& 86.8\% & 0.058 & 3.6\%		& 9.7\%	& 86.7\%  	    \\ \hline
Germany 			& 0.023 & 0.2\%	& 51.8\% 	& 48.0\% & 0.023 & 4.1\%		& 13.6\%	& 82.3\%  	    \\  \hline
Italy 				& 0.021 & 0.1\%	& 45.1\% 	& 54.8\% & 0.064 & 0.2\%		& 58.8\%	& 41.1\%    \\ \hline
Norway			& 0.023 & 16.1\%	& 0.01\% 	& 83.9\% & 0.093 & 15.5\%	& 41.0\%	& 43.5\%  	     \\ \hline
United Kingdom	& 0.023 & 14.0\%	& 14.81\% & 71.2\% & 0.069 & 83.4\%	& 1.2\%	& 15.4\%     \\ \hline
United States 		& 0.014 & 8.2\%	& 23.30\% & 68.5\% & 0.054 & 70.4\%	& 7.6\%	& 22.0\%       \\ \hline
Average 			& 0.026 & 9.5\%	& 29.0\% 	& 61.5\% & 0.058 & 27.9\%	& 31.0\%	& 41.1\% 	     \\ \hline
\end{tabular}
\caption{Mean squared error for simulated trajectories of the modified Goodwin model  - 1960 to 2010.}
\label{mse}
\end{table}

\begin{table}[H]
\centering
\resizebox{\textwidth}{!}{%
\begin{tabular}{|l|l|l|l|l|l|l|l|l|l|l|}
\hline
Country    & 			Australia   	& Canada      	& Denmark     	& Finland     	& France 		& Germany     	& Italy       	& Norway      	& UK 		& US 		 	\\ \hline
$\widehat{\log a_0}$   	& -$3.10^{+}$ 	& -$3.18^{+}$ 	& -$1.57^{+}$ 	& -$4.14^{+}$ 	& -$3.81^{+}$ 	& -$3.77^{+}$	& -$3.81^{+}$ 	& -$1.36^{+}$ 	& -$4.21^{+}$	& -$3.21^{+}$     \\ \hline
$\widehat\alpha$  		& $0.015^{+}$ 	& $0.013^{+}$ 	& $0.018^{+}$ 	& $0.029^{+}$ 	& $0.022^{+}$ 	& $0.027^{+}$	& $0.021^{+}$ 	& $0.023^{+}$ 	& $0.021^{+}$	& $0.016^{+}$	  \\ \hline
Rsquare    			& 0.988       	& 0.968       	& 0.976       	& 0.975       	& 0.906       	& 0.937		& 0.829       	& 0.978       	& 0.990          	& 0.986         	        \\ \hline
adjRsquare 			& 0.987       	& 0.967       	& 0.975       	& 0.975       	& 0.904       	& 0.935 		& 0.825       	& 0.978       	& 0.990          	& 0.985         	       \\ \hline
Fstat      				& $3,902^{+}$ 	& $1,462^{+}$ 	& $1,969^{+}$ 	& $1,932^{+}$ 	& $474^{+}$   	& $448^{+}$ 	& $237^{+}$   	& $2,174^{+}$ 	& $4,905^{+}$	& $3,370^{+}$	   \\ \hline
LBQstat    			& $37^{+}$    	& $85^{+}$    	& $74^{+}$    	& $99^{+}$    	& $138^{+}$   	& $76^{+}$	& $135^{+}$   	& $63^{+}$    	& $46^{+}$       & $69^{+}$	     \\ \hline
JBStat     				& 2.60        	& 1.29        	& $14.33^{+}$ 	& 5.59        	& 4.84        	& 2.56 		& 5.63        	& $17.52^{+}$ 	& 0.70           	& 1.91          	        \\ \hline
ARCHstat   			& $25.27^{+}$ 	& $35.25^{+}$ 	& $22.21^{+}$ 	& $29.19^{+}$ 	& $44.51^{+}$ 	& $21.97^{+}$ 	& $43.21^{+}$ 	& $39.70^{+}$ 	& $14.33^{+}$	& $16.23^{+}$	 \\ \hline
\end{tabular}
}
\caption[Productivity Growth]{Estimate of Productivity growth given by equation \eqref{productvtyeqn}. The symbol + indicates significance level of 1\%}
\label{productivityGrowth}
\end{table}

\begin{table}[H]
\centering
\resizebox{\textwidth}{!}{%
\begin{tabular}{|l|l|l|l|l|l|l|l|l|l|l|}
\hline
Country    			& Australia   	& Canada      	& Denmark     	& Finland     	& France       		& Germany	& Italy       	& Norway      	& UK 		& US 		 \\ \hline
$\widehat{\log N_0}$  	& $8.39^{+}$  	& $8.89^{+}$  	& $7.74^{+}$  	& $7.73^{+}$  	& $9.92^{+}$   		& $10.14^{+}$	& $9.92^{+}$  	& $7.32^{+}$  	& $10.07^{+}$  & $11.20^{+}$    \\ \hline
$\widehat\beta$  		& $0.020^{+}$ 	& $0.020^{+}$ 	& $0.006^{+}$ 	& $0.003^{+}$ 	& $0.008^{+}$  		& $0.006^{+}$	& $0.006^{+}$ 	& $0.011^{+}$ 	& $0.005^{+}$  & $0.016^{+}$    \\ \hline
Rsquare    			& 0.989       	& 0.965       	& 0.900       	& 0.786       	& 0.995        		& 0.822 		& 0.938       	& 0.978       	& 0.963          	& 0.979                \\ \hline
adjRsquare 			& 0.989       	& 0.964       	& 0.898       	& 0.782       	& 0.995        		& 0.816		& 0.937       	& 0.977       	& 0.963          	& 0.978                \\ \hline
Fstat      				& $4,470^{+}$ 	& $1,359^{+}$ 	& $443^{+}$   	& $180^{+}$   	& $10,774^{+}$		& $138^{+}$	& $746^{+}$   	& $2,134^{+}$ 	& $1,286^{+}$  & $2,248^{+}$      \\ \hline
LBQstat    			& $123^{+}$   	& $182^{+}$   	& $138^{+}$   	& $121^{+}$   	& $82^{+}$     		& $51^{+}$	& $65^{+}$    	& $104^{+}$   	& $81^{+}$       & $148^{+}$         \\ \hline
JBStat     				& 4.21        	& 4.45        	& 1.72        	& 0.92        	& 1.75         		& 1.33 		& $13.34^{+}$ 	& 2.88        	& 3.17           	& 1.76     	        \\ \hline
ARCHstat   			& $40.89^{+}$ 	& $43.32^{+}$ 	& $36.40^{+}$ 	& $38.52^{+}$ 	& $18.61^{+}$  		& $19.88^{+}$	& $39.11^{+}$ 	& $30.06^{+}$ 	& $24.15^{+}$  & $36.54^{+}$    \\ \hline
\end{tabular}
}
\caption[Labour Growth]{Estimate of Labour Force growth given by equation \eqref{labourgrowtheqn}. The symbol + indicates significance level of 1\%}
\label{laborGrowth}
\end{table}

\begin{table}[H]
\centering
\resizebox{\textwidth}{!}{%
\begin{tabular}{|c|c|c|c|c|c|}
\hline
Country & real wage growth & employment rate & productivity growth & inflation & nominal wage growth \\ \hline
Australia & 0.001 & 0.449 & 0.001 & 0.260 & 0.153 \\ \hline
Canada & 0.001 & 0.510 & 0.001 & 0.140 & 0.220 \\ \hline
Denmark & 0.404 & 0.535 & 0.001 & 0.410 & 0.200 \\ \hline
Finland & 0.001 & 0.073 & 0.001 & 0.160 & 0.424 \\ \hline
France & 0.063 & 0.655 & 0.050 & 0.695 & 0.675 \\ \hline
Germany & 0.147 & 0.432 & 0.046 & 0.157 & 0.256 \\ \hline
Italy & 0.073 & 0.341 & 0.013 & 0.607 & 0.530 \\ \hline
Norway & 0.001 & 0.549 & 0.002 & 0.001 & 0.114 \\ \hline
UK & 0.001 & 0.293 & 0.001 & 0.223 & 0.138 \\ \hline
US & 0.001 & 0.402 & 0.001 & 0.438 & 0.096 \\ \hline
\end{tabular}
}
\caption[Unit Root test]{p-values for augmented Dicky-Fuller (ADF) test.}
\label{unitRoot}
\end{table}

\begin{table}[H]
\centering
\begin{tabular}{|l|c|c|c|c|c|}
\hline
Country & lag 1 & lag 2 & lag 3 & lag 4 & lag 5 \\ \hline
Australia & 0.967 & 0.926 & 0.938 & 0.948 & 0.970 \\ \hline
Canada & 0.912 & 0.242 & 0.358 & 0.520 & 0.664 \\ \hline
Denmark & 0.742 & 0.946 & 0.279 & 0.334 & 0.463 \\ \hline
Finland & 0.714 & 0.841 & 0.432 & 0.594 & 0.732 \\ \hline
France & 0.555 & 0.838 & 0.859 & 0.508 & 0.453 \\ \hline
Germany & 0.795 & 0.603 & 0.786 & 0.525 & 0.282 \\ \hline
Italy & 0.313 & 0.594 & 0.719 & 0.827 & 0.872 \\ \hline
Norway & 0.940 & 0.935 & 0.795 & 0.846 & 0.922 \\ \hline
UK & 0.948 & 0.997 & 0.869 & 0.323 & 0.425 \\ \hline
US & 0.687 & 0.642 & 0.794 & 0.598 & 0.298 \\ \hline
\end{tabular}
\caption[Serial Correlation in Unrestricted ECM]{p-values for the alternative hypothesis that 
the errors in the unrestricted ECM given by equation \eqref{uecm1} are AR(m) for $m=1,\ldots,5$.}
\label{serialCorrel_modela}
\end{table}

\begin{table}[H]
\centering
\resizebox{\textwidth}{!}{%
\begin{tabular}{|c|c|c|c|c|c|c|c|c|c|c|}
\hline
Country & Australia & Canada & Denmark & Finland & France & Italy & Norway & UK & US & Germany \\ \hline
F statistics & 15.548 & 17.154 & 33.071 & 21.107 & 12.574 & 8.519 & 21.421 & 13.830 & 8.019 & 5.651 \\ \hline
\end{tabular}
}
\caption[F-test for restricted error correction model]{F-test for  $H_0: \phi_2 = \phi_3 = 0$ restriction in equation \eqref{uecm1}. Lower and upper bounds for $I(0)$ and $I(1)$ at the 1\%, 5\% and 10\% levels are [7.560, 8.685], [5.220, 6.070] and [4.190, 4.940], respectively.}
\label{ftest_modela}
\end{table}

\begin{table}[H]
\centering
\begin{tabular}{|c|c|c|c|c|c|c|c|c|c|}
\hline
Country & Variable & $\widehat\gamma$ & $\widehat\rho$ & AdjR2 & lag 1 & lag 2 & lag 3 & lag 4 & lag 5 \\ \hline
Australia & Coeff & -0.215 & 0.242 & 0.086 & & & & & \\ \cline{2-10} 
 & pValue & 0.031 & 0.022 & & 0.155 & 0.363 & 0.560 & 0.646 & 0.743 \\ \hline
Canada & Coeff & -0.095 & 0.115 & 0.010 & & & & & \\ \cline{2-10} 
 & pValue & 0.283 & 0.230 & & 0.407 & 0.173 & 0.196 & 0.307 & 0.438 \\ \hline
Denmark & Coeff & -0.330 & 0.367 & 0.216 & & & & & \\ \cline{2-10} 
 & pValue & 0.001 & 0.000 & & 0.397 & 0.298 & 0.061 & 0.109 & 0.181 \\ \hline
Finland & Coeff & -0.258 & 0.303 & 0.274 & & & & & \\ \cline{2-10} 
 & pValue & 0.000 & 0.000 & & 0.493 & 0.571 & 0.174 & 0.284 & 0.393 \\ \hline
France & Coeff & -0.491 & 0.549 & 0.755 & & & & & \\ \cline{2-10} 
 & pValue & 0.000 & 0.000 & & 0.090 & 0.237 & 0.406 & 0.210 & 0.311  \\ \hline
Germany & Coeff & -0.699 & 0.747 & 0.673  & & & & &\\ \cline{2-10} 
 & pValue & 0.000 & 0.000 & & 0.287 & 0.342 & 0.483 & 0.576 & 0.276  \\ \hline
Italy & Coeff & -0.891 & 0.982 & 0.507 & & & & & \\ \cline{2-10} 
 & pValue & 0.000 & 0.000 & & 0.015 & 0.010 & 0.011 & 0.021 & 0.035  \\ \hline
Norway & Coeff & -0.574 & 0.609 & 0.039  & & & & &\\ \cline{2-10} 
 & pValue & 0.100 & 0.090 & & 0.868 & 0.852 & 0.745 & 0.800 & 0.891 \\ \hline
UK & Coeff & -0.108 & 0.135 & 0.045  & & & & &\\ \cline{2-10} 
 & pValue & 0.131 & 0.076 & & 0.085 & 0.168 & 0.289 & 0.078 & 0.097 \\ \hline
US & Coeff & -0.227 & 0.257 & 0.086  & & & & &\\ \cline{2-10} 
 & pValue & 0.031 & 0.022 & & 0.057 & 0.134 & 0.224 & 0.358 & 0.322 \\ \hline
\end{tabular}
\caption{Long term estimates for Phillips curve given by equation \eqref{ltpc} and p-values for the alternative hypothesis that 
the errors are AR(m) for $m=1,\ldots,5$ in the serial correlation test.}
\label{linearPC}
\end{table}

\begin{table}[H]
\centering
\begin{tabular}{|l|c|c|c|c|c|}
\hline
Country & Variable & $\widehat\phi_{10}$ & $\widehat\phi_{11}$ & $\widehat\phi_{12}$ & AdjR2 \\ \hline
Australia & Coeff & 0.000 & 0.023 & -0.826 & 0.390 \\ \cline{2-6} 
 & pValue & 0.944 & 0.941 & 0.000 &  \\ \hline
Canada & Coeff & 0.000 & 0.034 & -0.873 & 0.425 \\ \cline{2-6} 
 & pValue & 0.859 & 0.884 & 0.000 &  \\ \hline
Denmark & Coeff & -0.001 & 0.343 & -1.097 & 0.586 \\ \cline{2-6} 
 & pValue & 0.654 & 0.176 & 0.000 &  \\ \hline
Finland & Coeff & 0.000 & 0.375 & -0.953 & 0.473 \\ \cline{2-6} 
 & pValue & 0.888 & 0.076 & 0.000 &  \\ \hline
France & Coeff & -0.001 & 0.103 & -0.679 & 0.330 \\ \cline{2-6} 
 & pValue & 0.436 & 0.663 & 0.000 &  \\ \hline
Germany & Coeff & -0.001 & 0.192 & -0.766 & 0.237 \\ \cline{2-6} 
 & pValue & 0.777 & 0.676 & 0.003 &  \\ \hline
Italy & Coeff & -0.001 & -0.263 & -0.580 & 0.245 \\ \cline{2-6} 
 & pValue & 0.754 & 0.570 & 0.000 &  \\ \hline
Norway & Coeff & 0.000 & 0.902 & -0.983 & 0.469 \\ \cline{2-6} 
 & pValue & 0.963 & 0.388 & 0.000 &  \\ \hline
UK & Coeff & 0.000 & 0.278 & -0.762 & 0.355 \\ \cline{2-6} 
 & pValue & 0.981 & 0.279 & 0.000 &  \\ \hline
US & Coeff & 0.000 & -0.249 & -0.583 & 0.279 \\ \cline{2-6} 
 & pValue & 0.841 & 0.175 & 0.000 &  \\ \hline
\end{tabular}
\caption{Restricted Error Correction Model \eqref{recm1} confirming a negative and significant coefficient $\hat\phi_{12}$ for the lagged error term $v_{t-1}$}
\label{RestrictedECM}
\end{table}

\end{document}